# About quantum mechanics interpretation

## Alexander G. Kyriakos

*Saint-Petersburg State Institute of Technology,
St.Petersburg, Russia*

Present address:
   *Athens, Greece*
*e-mail: lelekous@otenet.gr*

## Abstract

The quantum electron theory analysis is the object of this article. As it is known, the modern (Copenhagen's) interpretation of quantum mechanics is correct. However, it doesn't satisfy many physicists, who have the opinion that the modern quantum mechanics is a phenomenological theory. The suggested theory is the new quantum mechanics interpretation. It is entirely according to the modern interpretation, but it naturally explains the postulates of the modern quantum mechanics and gives many other results.

## Contents

# 1. Introduction

According to the Standard theory form, the elementary particles theory is the generalization of the quantum electrodynamics. The history of the question about the foundation of the electron theory is well known. The attempts to explain the electron structure and its properties were already undertaken in the last centuries (Lord Kelvin, H.A.Lorentz, Lois de Broglie et.al.) and have continued until today [1].

In the following article within the scope of quantum electrodynamics we show that, the electron arises by the photon electromagnetic field space transformation, which results to the electron spinor appearance.

# 2. Particles production hypotheses

Let's consider the particle-antiparticle production conditions.

One $\gamma$-quantum cannot turn spontaneously into the electron-positron pair, although it interacts with the electron-positron vacuum. For the pair production, at first, the following mass correlation is necessary: $\varepsilon_p \geq 2m_e c^2$ (where $\varepsilon_p$ is the photon energy, $m_e$ - the electron mass and $c$ - the light velocity). At second, the presence of the other particle, having the electromagnetic field, is needed. It can be some other $\gamma$-quantum, the electron $e^-, e^+$, an atom nucleus $Ze$ etc. For example, we have the typical reaction:

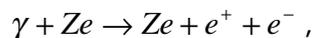
$$\gamma + Ze \rightarrow Ze + e^+ + e^-,$$

which means that, while moving through the particle field the photon(or may be virtual photon) takes some transformation, which corresponds to the pair production. Considering the fact that Pauli's matrices describe the vector space rotations and taking in account the optical-mechanical analogy analysis also, we can assume that the above transformation is a field distortion. From this follows the **distortion hypothesis:**

*By the fulfilment of the pair production conditions the distortion of the electromagnetic field of photon can take place; as result photon is able to move along the closed trajectory, making some stable construction named elementary particle.*

Also the following preposition, which is in accordance with the Heisenberg uncertainty principle, is necessary to the entirety of the theory: *all photons contain one wave period.*

# 3. Electrodynamics' form of quantum electron theory

If the object, formed after the wave division, is the electron-positron pair, it must be defined by spinors, which have other transformation properties than the vectors of Maxwell's electromagnetic field. In addition, as it is known, the modern theories of elementary particles are described by non-linear field equations.

Therefore, according to our suppositions *the Dirac equations must be some modification of Maxwell's equations and thereto the limit of some non-linear equations.*

Let us prove it.

### 3.1. Electrodynamics' form of Dirac's equation

*About the electron, as the simplest paticle, we can suppose that the photon trajectory is circular.*

Consider the linear photon, moving along $y$-axis. In a more general case it has the two possible polarisations and contains the field vectors $E_x, E_z, H_x, H_z$ $(E_y = H_y = 0)$. Such photon can form the ring only on the $(x,o,y)$ or the $(y,o,z)$ plains.

The bispinor form of Dirac's equations can be written as one equation [2]:

$$\hat{\varepsilon}\psi + c\hat{\vec{\alpha}} \; \hat{\vec{p}} + \hat{\beta} \; m_e c^2 \psi = 0, \qquad (3.1)$$

where $\hat{\vec{\alpha}}, \hat{\beta}$ - are Dirac's matrices, $\hat{\varepsilon} = i\hbar \dfrac{\partial}{\partial t}$, $\hat{\vec{p}} = -i\hbar \vec{\nabla}$ - the operators of energy and momentum, $\psi$ is the wave function named bispinor, which has four component.

Put the following semi-photon bispinor:

$$\psi = \begin{pmatrix} E_z \\ E_x \\ iH_z \\ iH_x \end{pmatrix} \qquad (3.2)$$

Taking into account that $\psi = \psi(y)$, from (4.1) we obtain:

$$\begin{cases} rot \; \vec{E} + \dfrac{1}{c} \dfrac{\partial \; \vec{H}}{\partial \; t} = i\dfrac{\omega}{c} \vec{H}, \\ rot \; \vec{H} - \dfrac{1}{c} \dfrac{\partial \; \vec{E}}{\partial \; t} = -i\dfrac{\omega}{c} \vec{E}, \end{cases} \qquad (3.3)$$

where $\omega = \dfrac{m_e c^2}{\hbar}$. The equations (3.3) are the Maxwell equations with current [3]. It is interesting that together with the electrical current the magnetic current also exists here. This current is equal to zero by Maxwell's theory, but its existence by Dirac doesn't contradict to the quantum theory. (According to the full theory the magnetic current appearance relates to the origin photon circular polarisation and integrally this current is equal to zero).

So, our first supposition is right. Now we must show that all the quantum mechanics can have the electrodynamics' form.

### 3.2. Electrodynamics' and quantum forms of electron-positron pair production theory

Using (3.2), we can write the equation of the electromagnetic wave moved along the any axis in form:

$$\left(\hat{\varepsilon}^2 - c^2 \hat{\vec{p}}^2\right)\psi = 0, \tag{3.4}$$

The equation (3.4) can also be written in the following form:

$$\left[\left(\hat{\alpha}_o \hat{\varepsilon}\right)^2 - c^2 \left(\hat{\vec{\alpha}} \ \hat{\vec{p}}\right)^2\right] \psi = 0, \tag{3.5}$$

where $\hat{\alpha}_o = \hat{1}$ is the unit matrix. In fact, taking into account that

$$\left(\hat{\alpha}_o \hat{\varepsilon}\right)^2 = \hat{\varepsilon}^2, \quad \left(\hat{\vec{\alpha}} \ \hat{\vec{p}}\right)^2 = \hat{\vec{p}}^2,$$

we see that equations (3.4) and (3.5) are equivalent.

Factorising (3.5) and multiplying it from left on Hermithian-conjugate function $\psi^+$ we get:

$$\psi^+ \left(\hat{\alpha}_o \hat{\varepsilon} - c\hat{\vec{\alpha}} \ \hat{\vec{p}}\right) \left(\hat{\alpha}_o \hat{\varepsilon} + c\hat{\vec{\alpha}} \ \hat{\vec{p}}\right) \psi = 0, \tag{3.6}$$

The equation (4.6) may be disintegrated on two equations:

$$\psi^+ \left(\hat{\alpha}_o \hat{\varepsilon} - c\hat{\vec{\alpha}} \ \hat{\vec{p}}\right) = 0, \tag{3.7}$$

$$\left(\hat{\alpha}_o \hat{\varepsilon} + c\hat{\vec{\alpha}} \ \hat{\vec{p}}\right) \psi = 0, \tag{3.8}$$

It is not difficult to show (using (3.2)) that the equations (3.7) and (3.8) are Maxwell's equations without current and, at the same time, are Dirac's electron-positron equations without mass.

*In accordance with our assumption, the reason for current appearance is the electromagnetic wave motion along a curvilinear trajectory.* We will show the appearance of the current, using the general methods of the distortion field investigation [4], but the same result can be obtained in the vector form (see Appendix, chapter A1.2.). The question is about the tangent space introduction at every point of the curvilinear space, in which the orthogonal axis system moves. This corresponds to the fact, that the wave motion along a circular trajectory is accompanied by the motion of the rectangular basis, built on vectors ($\vec{E}, \vec{S}, \vec{H}$), where $\vec{S}$ is the Poynting vector.

For the generalisation of Dirac's equation in Riemann's geometry it is necessary [4] to replace the usual derivative $\partial_\mu \equiv \partial / \partial x_\mu$ (where $x_\mu$ is the co-ordinates in the 4-space) with the covariant derivative: $D_\mu = \partial_\mu + \Gamma_\mu$ ($\mu = 0, 1, 2, 3$ are the summing indexes), where $\Gamma_\mu$ is the analogue of Christoffel's symbols in the case of the spinors theory. When a spinor moves along the beeline, all $\Gamma_\mu = 0$, and we have a usual derivative. But if a spinor moves along the curvilinear trajectory, then not all $\Gamma_\mu$ are equal to zero and a supplementary term appears. Typically, the last one is not the derivative, but is equal to the product of the spinor itself with some coefficient $\Gamma_\mu$. Thus we can assume that the supplementary term a longitudinal field is, i.e. it is a current. So from (3.7-3.8) we obtain:

$$\alpha^\mu D_\mu \psi = \alpha^\mu (\partial_\mu + \Gamma_\mu) \psi = 0$$

According to general theory [4] the increment in spinor $\Gamma_\mu$ has the form of the energy-momentum 4-vector. It is logical (see also

Appendix 1 below) to identify $\Gamma_\mu$ with 4-vector of energy-momentum of the own electron field:

$$\Gamma_\mu = \{\varepsilon_s, c\vec{p}_s\}, \qquad (3.9)$$

Then equations (3.7) and (3.8) in the curvilinear space will be have the view:

$$\psi^+ [\,(\hat{\alpha}_o \hat{\varepsilon} - c\hat{\vec{\alpha}}\,\hat{\vec{p}}) - (\hat{\alpha}_o \varepsilon_s - c\hat{\vec{\alpha}}\,\vec{p}_s)\,] = 0, \qquad (3.10)$$

$$[\,(\hat{\alpha}_o \hat{\varepsilon} + c\hat{\vec{\alpha}}\,\hat{\vec{p}}) + (\hat{\alpha}_o \varepsilon_s + c\hat{\vec{\alpha}}\,\vec{p}_s)\,]\,\psi = 0, \qquad (3.11)$$

According to the energy conservation law we can write:

$$\hat{\alpha}_o \varepsilon_s \pm c\hat{\vec{\alpha}}\,\vec{p}_s = \mp \hat{\beta}\, m_e c^2, \qquad (3.12)$$

Substituting (3.12) in (3.10) and (3.11) we will arrive at the usual kind of Dirac's equation with the mass:

$$\psi^+ [\,(\hat{\alpha}_o \hat{\varepsilon} - c\hat{\vec{\alpha}}\,\hat{\vec{p}}) - \hat{\beta}\, m_e c^2\,] = 0, \qquad (3.13)$$

$$[\,(\hat{\alpha}_o \hat{\varepsilon} + c\hat{\vec{\alpha}}\,\hat{\vec{p}}) + \hat{\beta}\, m_e c^2\,]\,\psi = 0, \qquad (3.14)$$

Therefore, our supposition is correct and *the electron theory can be represented as the spinning semi-photon theory.*

Note that *the above calculating procedure corresponds to the photon symmetry spontaneous breakdown and the mass appearance process*.

Note also: *because $E_y = H_y = 0$ is true by any transformation, Dirac's electron equation bispinor has only four components as special property.*

The objects that corespond to equations (3.13) and (3.14) we will name the spinning semi-photons (see Appendix 1).

### 3.3. Electrodynamics' sense of bilinear forms and of the statistical interpretation of the wave function

It is known that there are 16 Dirac's matrices of 4x4 dimensions. (We use the set of matrices, which originally Dirac himself used, and we will name it $\alpha$-set). Enumerate main Dirac's matrices [2]:

1) $\hat{\alpha}_4 \equiv \hat{\beta}$ is the scalar, 2) $\hat{\alpha}_\mu = \{\hat{\alpha}_0, \hat{\vec{\alpha}}\} \equiv \{\hat{\alpha}_0, \hat{\alpha}_1, \hat{\alpha}_2, \hat{\alpha}_3\}$ is the 4-vector,

3) $\hat{\alpha}_5 = \hat{\alpha}_1 \cdot \hat{\alpha}_2 \cdot \hat{\alpha}_3 \cdot \hat{\alpha}_4$ is the pseudoscalar.

Using (3.2) and taking in account that $\psi = \psi(y)$ and $\psi^+ = (E_z\ E_x\ -iH_z\ -iH_x)$ (where ($^+$) is the Hermithian conjugation sign) it is easy to obtain the electrodynamics' expression of bispinors, corresponding to these matrices:

1) $\psi^+ \hat{\alpha}_4 \psi = (E_x^2 + E_z^2) - (H_x^2 + H_z^2) = \vec{E}^2 - \vec{H}^2 = 8\pi\, I_1$, where $I_1$ is the first scalar of Maxwell's theory;

2) $\psi^+ \hat{\alpha}_o \psi = \vec{E}^2 + \vec{H}^2 = 8\pi\, U$ and $\psi^+ \hat{\alpha}_y \psi = 8\pi\, c\vec{g}_y$. In other words, the 4-vector bispinor value is the energy-momentum 4-vector $\left\{\dfrac{1}{c}U, \vec{g}\right\}$.

3) $\psi^+\hat{\alpha}_5\psi = 2(E_xH_x + E_zH_z) = 2(\vec{E}\cdot\vec{H})$ is the pseudoscalar of the electromagnetic field, which gives the second scalar of the electromagnetic field theory: $(\vec{E}\cdot\vec{H})^2 = I_2$.

From Dirac's equation the continuity equation for the probability can be obtained:

$$\frac{\partial P_{pr}(\vec{r},t)}{\partial t} + div\, \vec{S}_{pr}(\vec{r},t) = 0, \qquad (3.15)$$

Here $P_{pr}(\vec{r},t) = \psi^+\hat{\alpha}_0\psi$ is the probability density, and $\vec{S}_{pr}(\vec{r},t) = -c\psi^+\hat{\vec{\alpha}}\psi$ is the probability-flux density. Using the above results we can obtain: $P_{pr}(\vec{r},t) = 8\pi\, U$ and $\vec{S}_{pr} = c^2\vec{g} = 8\pi\, \vec{S}$. Taking in account the above results from (3.15) we get the equation:

$$\frac{\partial U}{\partial t} + div\, \vec{S} = 0, \qquad (3.16)$$

which is the energy conservation law of the electron electromagnetic field.

*So, the non-normalised probability density and probability-flux density are the energy density and energy flux density of the electromagnetic field, accordingly.* Normalising to unit the energy distribution we come to the distribution of probability density of the one and other value. For free immobile electron the normalisation means: $\int_0^\infty U d\tau = m_e c^2$, or $\frac{1}{8\pi\, m_e c^2}\int \overline{P}_{pr} d\tau = 1$, where $\overline{P}_{pr}$ is non-normalised probability density. Using the change: $\psi \to \sqrt{8\pi\, m_e c^2}\,\psi'$, where $\psi'$ is the normalised wave function, from the last equation we have correlation $\int \psi'^+\psi' = 1$, which coincides with quantum normalisation.

It is not difficult to understand now the method of the quantum values calculations as the operators' eigenvalues: *if the elementary particles are the spinning electromagnetic waves, their interaction is the wave interference. In this case the calculation of probability density is the calculation of energy maximums of interference. As it is well known, in classical physics such calculation is the eigenvalue problem.*

### 3.4. Lorentz's force

The expression of Lorentz's force by the energy-momentum tensor of electromagnetic field $\tau_\mu^\nu$ is well known [3]:

$$f_{\mu\nu} = -\frac{1}{4\pi}\frac{\partial \tau_\mu^\nu}{\partial x^\nu} \equiv -\frac{1}{4\pi}\partial_\nu \tau_\mu^\nu, \qquad (3.17)$$

This tensor is symmetrical and has the following components:

$$\tau_{ij} = -(E_iE_j + H_iH_j) + \frac{1}{2}\delta_{ij}(\vec{E}^2 + \vec{H}^2),$$

$$\tau_{i0} = 4\pi\, S_i = [\vec{E}\times\vec{H}]_i = \frac{4\pi}{c}(\vec{S})_i, \qquad (3.18)$$

$$\tau_{00} = 4\pi\, U = \frac{1}{2}(\vec{E}^2 + \vec{H}^2),$$

where $i, j = 1,2,3$, and $\delta_{ij} = 0$ when $i = j$, and $\delta_{ij} = 1$ when $i \neq j$.
Using (3.18) we can write:

$$f_1 = f_3 = 0$$

$$f_2 \equiv -\left(\frac{\partial \vec{g}}{\partial t} + grad\ U\right), \quad (3.19)$$

$$f_0 = -\left(\frac{1}{c}\frac{\partial U}{\partial t} + c\ div\ \vec{g}\right). \quad (3.20)$$

In the general case $\vec{g} = g\ \vec{\tau}$ and $\dfrac{\partial \vec{g}}{\partial t} = \dfrac{\partial g}{\partial t}\vec{\tau} + g\dfrac{\partial \vec{\tau}}{\partial t}$. It's obvious that the supplement forces don't appear in the linear photon. In case if photon rolls up around any of the axis, which is perpendicular to the $y$-axis, we obtain the additional terms:

$$g_y \frac{\partial \vec{\tau}}{\partial t} = g_y \frac{v_s}{r_s}\vec{n} = g_y \omega_s \vec{n}, \quad (3.21)$$

Using equations (3.19) and (3.20) for the spinning photon $(E_x, H_z)$ we take the normal force components:

$$^{oz}f_2 = \frac{1}{4\pi}\frac{\omega_s}{c}E_x H_z \vec{n} = \frac{1}{c}j_\tau \cdot H_z \vec{n} \quad (3.22)$$

$$^{oz}f_0 = \frac{1}{4\pi}\frac{\omega_s}{c}E_x^2 = \frac{1}{c}j_\tau \cdot E_x, \quad (3.23)$$

for spinning photon $(E_z, H_x)$ we taken:

$$^{ox}f_2 = -\frac{1}{4\pi}\frac{\omega_s}{c}E_z H_x \vec{n} = -\frac{1}{c}j_\tau \cdot H_x \vec{n}, \quad (3.24)$$

$$^{ox}f_0 = -\frac{1}{4\pi}\frac{\omega_s}{c}E_z^2 = -\frac{1}{c}j_\tau \cdot E_z, \quad (3.25)$$

(the upper left index shows the spinning axis: $oz$ or $ox$).
As we see the Lorenz's force direction doesn't coincide with the $y$-axis, but is perpendicularly to it.

### 3.5. The electrodynamics' form of electron theory Lagrangian

As it is known [3,5], the Lagrangian of the free field Maxwell's theory is:

$$L_M = \frac{1}{8\pi}\left(\vec{E}^2 - \vec{H}^2\right), \quad (3.26)$$

As Lagrangian of Dirac's theory can be taken the expression [2]:

$$L_D = \psi^+\left(\hat{\varepsilon} + c\hat{\vec{\alpha}}\ \hat{\vec{p}} + \hat{\beta}\ m_e c^2\right)\psi, \quad (3.27)$$

For the wave moving along the $y$-axis the equation (3.27) can be written:

$$L_D = \frac{1}{c}\psi^+\frac{\partial \psi}{\partial t} - \psi^+\hat{\alpha}_y\frac{\partial \psi}{\partial y} - i\frac{m_e c}{\hbar}\psi^+\hat{\beta}\ \psi, \quad (3.28)$$

Transferring the each term of (3.28) in electrodynamics' form we obtain for the semi-photon equation Lagrangian:

$$L_s = \frac{\partial U}{\partial t} + div\ \vec{S} - i\frac{\omega_s}{8\pi}\left(\vec{E}^2 - \vec{H}^2\right), \quad (3.29)$$

where $\omega_s = \frac{2mc^2}{\hbar}$ (note that we must differ the complex conjugate field vectors $\vec{E}^*, \vec{H}^*$ and $\vec{E}, \vec{H}$).

The equation (3.29) can be written in other form. Using electrical and magnetic currents

$$j_\tau^e = i\frac{\omega_s}{4\pi}\vec{E} \quad \text{and} \quad j_\tau^m = i\frac{\omega_s}{4\pi}\vec{H}, \quad (3.30)$$

we take:

$$L_s = \frac{\partial U}{\partial t} + div\ \vec{S} - \frac{1}{2}\left(\vec{j}_\tau^e \vec{E} - \vec{j}_\tau^m \vec{H}\right), \quad (3.31)$$

Since $L_s = 0$ thanks to (3.1) we take the equation:

$$\frac{\partial U}{\partial t} + div\ \vec{S} - \frac{1}{2}\left(\vec{j}_\tau^e \vec{E} - \vec{j}_\tau^m \vec{H}\right) = 0, \quad (3.32)$$

which is the energy-momentum conservation law for the Maxwell equation with current.

According to (3.29) for the own electron electromagnetic wave the other kind of Maxwell's equation Lagrangian exists, which differs from (3.26):

$$L_M = \frac{1}{8\pi}\left(\vec{E}^2 - \vec{H}^2\right) = \frac{i}{\omega_s}\left(\frac{\partial U}{\partial t} + div\ \vec{S}\right), \quad (3.33)$$

## 3.6. Electromagnetic form of the free electron equation solution

From the general point of view for the $y$-direction photon two solutions must exist (don't taking in account the field signs):
1) for the wave, rotating around the $oz$-axis

$$^{oz}\psi = \begin{pmatrix} \psi_1 \\ 0 \\ 0 \\ \psi_4 \end{pmatrix} = \begin{pmatrix} E_z \\ 0 \\ 0 \\ iH_x \end{pmatrix}, \quad (3.34)$$

and 2) for the wave, rotating around the $ox$-axis

$$^{ox}\psi = \begin{pmatrix} 0 \\ \psi_2 \\ \psi_3 \\ 0 \end{pmatrix} = \begin{pmatrix} 0 \\ E_x \\ iH_z \\ 0 \end{pmatrix}, \quad (3.35)$$

Consider now the results of the exact theory. It's known [2] that the solution of Dirac's free electron equation (3.1) is the plane wave:

$$\psi_j = b_j e^{i(\vec{k}\vec{r} - \omega t + \phi)}, \quad j = 1, 2, 3, 4, \quad (3.36)$$

where the amplitudes $b_j$ are the numbers and $\phi$ is the initial wave phase. The functions (3.36) are the eigenfunctions for the energy-momentum operators, where $\hbar\omega$ and $\hbar\vec{k}$ are the energy-momentum eigenvalues. If we put (3.36) in (3.1) we obtain [2] the algebraic equation system, which are the linear equations for the $B_j = b_j e^{i\phi}$ values:

$$\begin{cases} (\varepsilon + m_e c^2)B_1 + cp_z B_3 + c(p_x - ip_y)B_4 = 0, \\ (\varepsilon + m_e c^2)B_2 + c(p_x + ip_y)B_3 - cp_z B_4 = 0, \\ (\varepsilon - m_e c^2)B_3 + cp_z B_1 + c(p_x - ip_y)B_2 = 0, \\ (\varepsilon - m_e c^2)B_4 + c(p_x + ip_y)B_1 - cp_z B_2 = 0, \end{cases} \quad (3.37)$$

where the energy $\varepsilon = \hbar\omega$ and the momentum $\vec{p} = \hbar\vec{k}$ are not the operators. This system has a solution only when the equation determinant is equal to zero:

$$\left(\varepsilon^2 - m_e^2 c^4 - c^2 \vec{p}^2\right)^2 = 0$$

Here for each $\vec{p}$, the energy $\varepsilon$ has either positive value $\varepsilon_+ = +\left(c^2 \vec{p}^2 + m_e^2 c^4\right)^{\frac{1}{2}}$ or negative value $\varepsilon_- = -\left(c^2 \vec{p}^2 - m_e^2 c^4\right)^{\frac{1}{2}}$.

So we have two linear-independent set of four orthogonal normalising spinors for $\varepsilon_+$:

1) $B_1 = -\dfrac{cp_z}{\varepsilon_+ + m_e c^2}, \quad B_2 = -\dfrac{c(p_x + ip_y)}{\varepsilon_+ + m_e c^2}, \quad B_3 = 1, \quad B_4 = 0,$ (3.38)

2) $B_1 = -\dfrac{c(p_x - ip_y)}{\varepsilon_+ + m_e c^2}, \quad B_2 = \dfrac{cp_z}{\varepsilon_+ + m_e c^2}, \quad B_3 = 0, \quad B_4 = 1,$ (3.39)

and for $\varepsilon_-$:

3) $B_1 = 1, \quad B_2 = 0, \quad B_3 = \dfrac{cp_z}{-\varepsilon_- + m_e c^2}, \quad B_4 = \dfrac{c(p_x + ip_y)}{-\varepsilon_- + m_e c^2},$ (3.40)

4) $B_1 = 0, \quad B_2 = 1, \quad B_3 = \dfrac{c(p_x - ip_y)}{-\varepsilon_- + m_e c^2}, \quad B_4 = -\dfrac{cp_z}{-\varepsilon_- + m_e c^2},$ (3.41)

Make analysis of these solutions.

**At first**, the existing of two linear independent solutions corresponds with two independent orientation of the electromagnetic wave vectors (3.34) and (3.35) and gives the unique logic explanation for this fact.

**Secondly**, since $\psi = \psi(y)$, we have $p_x = p_z = 0$, $p_y = m_e c$ and for the field vectors we obtain: from equations (3.38) and (3.39) for "positive" energy

$$B_+^{(1)} = \begin{pmatrix} 0 \\ b_2 \\ b_3 \\ 0 \end{pmatrix} \cdot e^{i\phi}, \quad B_+^{(2)} = \begin{pmatrix} b_1 \\ 0 \\ 0 \\ b_4 \end{pmatrix} \cdot e^{i\phi}, \quad (3.42)$$

and from (3.40) and (3.41) for "negative" energy

$$B_-^{(1)} = \begin{pmatrix} b_1 \\ 0 \\ 0 \\ b_4 \end{pmatrix} \cdot e^{i\phi}, \quad B_-^{(2)} = \begin{pmatrix} 0 \\ b_2 \\ b_3 \\ 0 \end{pmatrix} \cdot e^{i\phi} \quad , \tag{3.43}$$

which correspond to (3.34) and (3.35).

**At third**, it is not difficult to calculate the components value of the field vectors, putting $\varepsilon_+ = m_e c^2$ and $\phi = \frac{\pi}{2}$:

$$B_+^{(1)} = \begin{pmatrix} 0 \\ \frac{1}{2} \\ i \cdot 1 \\ 0 \end{pmatrix}, \quad B_+^{(2)} = \begin{pmatrix} -\frac{1}{2} \\ 0 \\ 0 \\ i \cdot 1 \end{pmatrix}, \tag{3.44}$$

And also for $\varepsilon_- = -m_e c^2$:

$$B_-^{(1)} = \begin{pmatrix} i \cdot 1 \\ 0 \\ 0 \\ -\frac{1}{2} \end{pmatrix}, \quad B_-^{(2)} = \begin{pmatrix} 0 \\ i \cdot 1 \\ \frac{1}{2} \\ 0 \end{pmatrix}, \tag{3.45}$$

As we can see, the above solutions correspond to the each half-period of the electromagnetic waves with $y$- direction. *The imaginary unit in the solutions indicates that the field vectors $\vec{E}$ and $\vec{H}$ are mutual orthogonal.*

It is interesting that between the $\vec{E}$ and $\vec{H}$ own field vectors of the semi-photon the definite correlation exists: $E_s = 2H_s$. We don't know the interpretation of this fact.

## 4. Non-linear equations and theirs Lagrangians

### 4.1. Non-linear equations of spinning semi-photon

The stability of the circular photon is possible only by the photons parts self-action. Obviously, the equation that describes the spinning semi-photon structure must be non-linear and like the equation, which describes photon-photon interaction [8,9].

We could introduce the self-field interaction to the spinning semi-photon equation like the external field is introduced to the quantum [4,5] and classical [6] field equations (putting herewith the photon mass equal to zero). But this equation may be obtained more rigorously.

Considering the electron as circular semi-photon we can write the energy conservation equation:

$$\varepsilon_s^2 = c^2 \vec{p}_s^{\,2} + m_e^2 c^4, \tag{4.1}$$

Then from (4.1) the linear conservation law can be written as:

$$\varepsilon_s = \pm \left( c\hat{\vec{\alpha}} \, \vec{p}_s + \hat{\beta} \, m_e c^2 \right), \tag{4.2}$$

The relation (4.2) is right, like as (4.1), since the second degree of (4.2) gives (4.1).

According to relationship (4.2) for the spinning semi-photon we can find:

$$\hat{\beta} m_e c^2 = -\left(\varepsilon_s - c\hat{\vec{\alpha}} \vec{p}_s\right),$$

Substitute (4.2) in Dirac's electron equation we obtain the non-linear integral equation:

$$\left[\hat{\alpha}_0 (\hat{\varepsilon} - \varepsilon_s) + c\hat{\vec{\alpha}} (\hat{\vec{p}} - \vec{p}_s)\right] \psi = 0, \qquad (4.3)$$

which is, as we propose, the common form of the spinning semi-photon equation.

In the electromagnetic form we have:

$$\varepsilon_s = \int_{\Delta\tau} U \, d\tau = \frac{1}{8\pi} \int_{\Delta\tau} \left(\vec{E}^2 + \vec{H}^2\right) d\tau, \qquad (4.4)$$

$$\vec{p}_s = \int_{\Delta\tau} \vec{g} \, d\tau = \frac{1}{c^2} \int_{\Delta\tau} \vec{S} \, d\tau = \frac{1}{4\pi} \int_{\Delta\tau} \left[\vec{E} \times \vec{H}\right] d\tau, \qquad (4.5)$$

where in general case $\Delta\tau$ have values from zero to infinity.

Let's consider the approximate form of this equation. Since the main part of the electron energy consists in the semi-photon volume, we can write:

$$\Delta\tau \cong \Delta\tau_s = 2\pi^2 r_s^3, \qquad (4.6)$$

Using the quantum form of $U$ and $\vec{S}$:

$$U = \frac{1}{8\pi} \psi^+ \hat{\alpha}_0 \psi, \qquad (4.7)$$

$$\vec{S} = -\frac{c}{8\pi} \psi^+ \hat{\vec{\alpha}} \psi = c^2 \vec{g}, \qquad (4.8)$$

and taking in to account that the free electron Dirac's equation solution is the plane wave:

$$\psi = \psi_0 \, e^{i(\omega t - ky)}, \qquad (4.9)$$

we can write (4.7) and (4.8) in the next approximately form:

$$\varepsilon_s = U \, \Delta\tau_s = \frac{\Delta\tau_s}{8\pi} \psi^+ \hat{\alpha}_0 \psi, \qquad (4.10)$$

$$\vec{p}_s = \vec{g} \, \Delta\tau_s = \frac{1}{c^2} \vec{S} \, \Delta\tau_s = -\frac{\Delta\tau_s}{8\pi c} \psi^+ \hat{\vec{\alpha}} \psi, \qquad (4.11)$$

Substitute (4.10) and (4.11) into (4.3) and taking in to account (4.6), we obtain the following approximate equation:

$$\frac{\partial \psi}{\partial t} - c\hat{\vec{\alpha}} \vec{\nabla} \psi + i \frac{1}{2\alpha_q c} \cdot r_s^3 \left(\psi^+ \hat{\alpha}_0 \psi - \hat{\vec{\alpha}} \psi^+ \hat{\vec{\alpha}} \psi\right) \psi = 0, \qquad (4.12)$$

It is not difficult to see that the equation (4.12) is a non-linear equation of the same type, which was investigated by Heisenberg e. al. [4,7] and which played for a while the role of the unitary field theory equation. In contrary to the last one, equation (5.12) is obtained logically correctly. Also the self-action constant $r_s$ [4,7] appears in (5.12) automatically and expresses the radius of the «bare» electron.

## 4.2. Equation of spinning semi-photon matter motion

As it is known [2,5], the motion equations can be found from the next operator equation:

$$\frac{d\hat{L}}{dt} = \frac{\partial \hat{L}}{\partial t} + \frac{1}{i\hbar}\left(\hat{L}\hat{H} - \hat{H}\hat{L}\right), \tag{4.13}$$

where $\hat{L}$ is the physical value operator, whose variation we want to find, and $\hat{H}$ is the Hamilton operator.

The Hamilton operator of the electron equation is equal:

$$\hat{H} = -c\hat{\vec{\alpha}}\,\hat{\vec{P}} - \hat{\beta}\,m_e c^2 + \varepsilon_e, \tag{4.14}$$

where $\vec{P}$ is the full momentum of electron.

Let us $\hat{\vec{P}}_s$ be the full momentum of the spinning semi-photon. For $\hat{L} = \vec{P}_s$ from (4.14) substituting $\vec{v}_s = c\hat{\vec{\alpha}}$, where $\vec{v}_s$ - velocity of the semi-photon matter, we obtain:

$$\frac{d\vec{P}_s}{dt} = \left(\frac{\partial \vec{p}_s}{\partial t} + grad\ \varepsilon_s\right) - \left[\vec{v}_s \times rot\ \vec{p}_s\right], \tag{4.15}$$

Since for the motionless electron $\frac{d\hat{\vec{P}}_s}{dt} = 0$, the motion equation is:

$$\left(\frac{\partial \vec{p}_s}{\partial t} + grad\ \varepsilon_s\right) - \left[\vec{v}_s \times rot\ \vec{p}_s\right] = 0, \tag{4.16}$$

Passing to the approximate value of the energy and momentum densities

$$\vec{g}_s = \frac{1}{\Delta\tau_s}\vec{p}_s, \quad U_s = \frac{1}{\Delta\tau_s}\varepsilon_s, \tag{5.17}$$

we obtain the equation of matter motion of spinning semi-photon:

$$\left(\frac{\partial \vec{g}_s}{\partial t} + grad\ U_s\right) - \left[\vec{v}_s \times rot\ \vec{g}_s\right] = 0 \tag{4.18}$$

Let's consider the motion equation of the ideal liquid in form of Lamb equation [8]. In case the external forces are absent, this equation has the form:

$$\left(\frac{\partial \vec{g}_l}{\partial t} + grad\ U_l\right) - \left[\vec{v}_l \times rot\ \vec{g}_l\right] = 0, \tag{4.19}$$

where $U_l, \vec{g}_l$ and $v_l$ are the energy, momentum density and velocity of ideal liquid accordingly.

Comparing (4.18) and (4.19) we see their mathematical identity.
According to (3.19) we have

$$\frac{\partial \vec{g}_s}{\partial t} + grad\ U_s = \vec{f}_L, \tag{4.20}$$

where $f_L$ is the Lorenz force. As it is known the term $\left[\vec{v}_l \times rot\ \vec{g}_l\right]$ in (4.19) is responsible for centripetal acceleration. Probably, we have the same in (4.18). If the "photon liquid" moves along the ring of $r_s$ radius, then the angular motion velocity $\omega_s$ is tied with $rot\ \vec{v}_s$ by expression:

$$rot\ \vec{v}_s = 2\vec{\omega}_s = 2\omega_s \vec{e}_z^{\,o}, \qquad (4.21)$$

and centripetal acceleration is:

$$\vec{a}_r = \frac{1}{2}\vec{v}_s \times rot\ \vec{v}_s = \frac{v^2}{r_s}\vec{e}_r^{\,o} = c\omega_s \vec{e}_r^{\,o}, \qquad (4.22)$$

where $\vec{e}_r^{\,o}$ is unit radius vector, $\vec{e}_z^{\,o}$ - is unit vector of $OZ$-axis. As a result the equation (4.21) has the form of Newton's law:

$$\rho_m \vec{a}_r = \vec{f}_L, \qquad (4.23)$$

The above calculations can be considered as a demonstration of the Erenfest theorem.

### 4.3. Lagrangian of non-linear semi-photon equation

The Lagrangian of non-linear equation is not difficult to obtain from Lagrangian of the linear Dirac's equation:

$$L_D = \psi^+ \left( \hat{\varepsilon} + c\hat{\vec{\alpha}}\ \hat{\vec{p}} + \hat{\beta}\ m_e c^2 \right)\psi, \qquad (4.24)$$

using the method by which we find the first degree non-linear equation. Substitute (4.3) into (4.24) we obtain:

$$L_N = \psi^+ \left( \hat{\varepsilon} - c\hat{\vec{\alpha}}\ \hat{\vec{p}} \right)\psi + \psi^+ \left( \varepsilon_s - c\hat{\vec{\alpha}}\ \vec{p}_s \right)\psi, \qquad (4.25)$$

The expression (4.25) represents the common form of Lagrangian of non-linear semi-photon equation.

Using (4.10) and (4.11) we can represent (4.25) in the approximate quantum form:

$$L_N = i\hbar \left[ \frac{\partial}{\partial t}\left[ \frac{1}{2}(\psi^+\psi) \right] - c\ div(\psi^+\hat{\vec{\alpha}}\psi) \right] + \frac{\Delta\tau_s}{8\pi}\left[ (\psi^+\psi)^2 - (\psi^+\hat{\vec{\alpha}}\psi)^2 \right], \qquad (4.26)$$

Normalising $\psi$-function we obtain:

$$L_N = \frac{1}{8\pi\ m_e c} L_N, \qquad (4.27)$$

Transform (4.25) in the electrodynamics' form, using equations (5.4) and (4.5), we find:

$$L_N = i\frac{\hbar}{2m_e c^2}\left( \frac{\partial U}{\partial t} + div\ \vec{S} \right) + \frac{\Delta\tau}{m_e c^2}\left( U^2 - c^2 \vec{g}^2 \right), \qquad (4.28)$$

Here accordingly to (3.29) the first summand may be replaced through

$$i\frac{\hbar}{2mc^2}\left( \frac{\partial U}{\partial t} + div\ \vec{S} \right) = \frac{1}{8\pi}\left( \vec{E}^2 - \vec{H}^2 \right), \qquad (4.29)$$

It is no difficult to transform the second summand, using the known electrodynamics' transformation:

$$(8\pi)^2 \left( U^2 - c^2 \vec{g}^2 \right) = \left( \vec{E}^2 + \vec{H}^2 \right)^2 - 4\left( \vec{E} \times \vec{H} \right)^2 = \left( \vec{E}^2 - \vec{H}^2 \right)^2 + 4\left( \vec{E} \cdot \vec{H} \right)^2, \qquad (4.30)$$

So we have:

$$L_N = \frac{1}{8\pi}\left( \vec{E}^2 - \vec{H}^2 \right) + \frac{\Delta\tau}{(8\pi)^2 m_e c^2}\left[ \left( \vec{E}^2 - \vec{H}^2 \right)^2 + 4\left( \vec{E} \cdot \vec{H} \right)^2 \right], \qquad (4.31)$$

As we see, the Lagrangian of non-linear spinning semi-photon equation contains only the invariant of Maxwell's theory. Accordingly to our conception the equation (4.31) must be Lagrangian of the semi-photon itself interaction. Really in the case of low changing fields can be calculated the Lagrangian of the photon-photon interaction [4,5]:

$$L_{p-p} = \frac{1}{8\pi}\left(\vec{E}^2 - \vec{H}^2\right) + b\left[\left(\vec{E}^2 - \vec{H}^2\right)^2 + 7\left(\vec{E}\cdot\vec{H}\right)^2\right] + ..., \qquad (4.32)$$

where $b = \frac{2}{45}\frac{e^4\hbar}{m_e^4 c^7}$, which coincides with (4.32) up to the number coefficients. This confirm our hypothesis that electron is an interacting photon itself.

Let's now analyse the quantum form of Lagrangian (4.31). The equation (4.25) can be written in form:

$$L_N = \psi^+\hat{\alpha}_\mu\partial_\mu\psi + \frac{\Delta\tau_s}{8\pi}\left[\left(\psi^+\hat{\alpha}_0\psi\right)^2 - \left(\psi^+\hat{\vec{\alpha}}\,\psi\right)^2\right], \qquad (4.33)$$

It is not difficult to see that *the electrodynamics' correlation (4.30) in quantum form has the known view of Fierz's correlation:*

$$\left(\psi^+\hat{\alpha}_0\psi\right)^2 - \left(\psi^+\hat{\vec{\alpha}}\,\psi\right)^2 = \left(\psi^+\hat{\alpha}_4\psi\right)^2 + \left(\psi^+\hat{\alpha}_5\psi\right)^2, \qquad (4.34)$$

Using (4.34) from (4.33) we obtain:

$$L_N = \psi^+\hat{\alpha}_\mu\partial_\mu\psi + \frac{\Delta\tau_s}{8\pi}\left[\left(\psi^+\hat{\alpha}_4\psi\right)^2 - \left(\psi^+\hat{\alpha}_5\psi\right)^2\right], \qquad (4.35)$$

The Lagrangian (4.35) practically (if to use the $\gamma$-set of Dirac's matrices instead of $\alpha$-set) coincide with the Nambu's and Jona-Losinio's Lagrangian [9], which is the Lagrangian of the relativistic superconductivity theory. As it is known this Lagrangian gives the solution to the problem of the appearance of the elementary particles mass by the mechanism of the vacuum symmetry spontaneous breakdown. It corresponds to the Cooper's pair decay on the electron and «hole» in the superconductivity theory.

## Appendix A1. Electromagnetic model of electron

From the above theory it follows that the electron is "composed" from electromagnetic field and therefore its structure can not be detected in the electromagnetic interactions. In modern physics this fact allows to consider an electron as a point particle, but it doesn't contradict to the existing of the electromagnetic electron structure.

### A1.1. Structure and parameters of the electron model

According to the hypotheses we can represent the electron as a torus with a radius $r_p$. We suppose also that the cross-section field radius is equal to $r_p$ (fig.1). It's obvious that the torus radius is equal to $r_p = \frac{\lambda_p}{2\pi}$, where $\lambda_p$ is the photon's wavelength.

In our case the photon characteristics are defined by the electron-positron pair production conditions: the photon energy $\varepsilon_p = 2m_e c^2$ and the circular frequency $\omega_p = \dfrac{\varepsilon_p}{\hbar} = \dfrac{2m_e c^2}{\hbar}$. Therefore, the photon wavelength is $\lambda_p = \dfrac{2\pi c}{\omega_p} = \dfrac{\pi \hbar}{m_e c}$, and the radius of torus (i.e. of the circular photon) is $r_p = \dfrac{\hbar}{2m_e c}$ (note that below the index "$p$" refers to the circular photon).

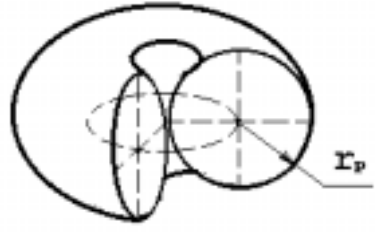

Fig.1. Electron model

**A1.2. Ring current**

Let the plane-polarised photon, which have the field vectors $(E_z, H_x)$ roll up with radius $r_p$ in the plane $(x',o',y')$ of a fixed co-ordinate system $(x',y',z',o')$, so that $E_z$ is in parallel to the plane $(x',o',y')$ and $H_x$ is perpendicularly to it. It could be said that the rectangular axes system $\{E_z, S_y, H_x\}$ moves along the tangent to the circumference, where $\vec{S}_y = [\vec{E} \times \vec{H}]_y$ is the $y$-component of the Poynting vector.

Let us show that due to the photon electromagnetic wave distortion the displacement ring current arises.

According to Maxwell [3] the displacement current is defined by the equation:

$$j_{dis} = \dfrac{1}{4\pi} \dfrac{\partial \vec{E}}{\partial t}, \qquad (A1.1)$$

The electrical field vector $\vec{E}$, which moves along the curvilinear trajectory (let it have the direction from the centre), can be written in form:

$$\vec{E} = -E \cdot \vec{n}, \qquad (A1.2)$$

where $E = |\vec{E}|$ and $\vec{n}$ is the normal unit-vector of the curve (having direction to the centre). The derivative of $\vec{E}$ with respect to $t$ can be represented as:

$$\frac{\partial \vec{E}}{\partial t} = -\frac{\partial E}{\partial t}\vec{n} - E\frac{\partial \vec{n}}{\partial t}, \qquad (A1.3)$$

Here the first term has the same direction as $\vec{E}$. The existence of the second term shows that at the wave distortion the supplementary displacement current appears. It is not difficult to show that it has a direction, tangential to the ring:

$$\frac{\partial \vec{n}}{\partial t} = -\frac{v_p}{r_p}\vec{\tau}, \qquad (A1.4)$$

where $\vec{\tau}$ is the tangential unit-vector, $v_p \equiv c$ is the photon velocity. Then the displacement current of the ring wave can be written:

$$\vec{j}_{dis} = -\frac{1}{4\pi}\frac{\partial E}{\partial t}\vec{n} + \frac{1}{4\pi}\omega_p E \cdot \vec{\tau}, \qquad (A1.5)$$

where $\omega_p = \dfrac{c}{r_p}$ is the angular velocity (or angular frequency), $\vec{j}_n = \dfrac{1}{4\pi}\dfrac{\partial E}{\partial t}\vec{n}$ and $\vec{j}_\tau = \dfrac{\omega_p}{4\pi}E \cdot \vec{\tau}$ are the normal and tangent components of the spinning photon current accordingly. So:

$$\vec{j}_{dis} = \vec{j}_n + \vec{j}_\tau, \qquad (A1.6)$$

The currents $\vec{j}_n$ and $\vec{j}_\tau$ are always mutually perpendicular, so that we can write in complex form:

$$j_{dis} = j_n + ij_\tau,$$

where $i = \sqrt{-1}$. From the above we can assume that *the appearance of an imaginary unity in the quantum mechanics is tied with the tangent current appearance.*

The same result can be obtained by using the general methods of the distortion field investigation (see chapter 3.2.).

## A1.3. Charge appearance and division hypothesis

We accept that the fields of the model particle lie within the torus volume. It's not difficult to calculate now the charge density of the spinning photon:

$$\rho_p = \frac{j_\tau}{c} = \frac{1}{4\pi}\frac{\omega_p}{c}E = \frac{1}{4\pi}\frac{1}{r_p}E, \qquad (A1.7)$$

The full charge of the particle can be defined by integrating by all the torus (circular photon) volume $\Delta\tau_p$:

$$q = \int_{\Delta\tau_p} \rho_p d\tau, \qquad (A1.8)$$

Using the model (fig.1) and taking $\vec{E} = \vec{E}(l)$, where $l$ is the length of the way along the circumference, we obtain:

$$q = \int\limits_{S_{tr}}\int\limits_0^{\lambda_p} \frac{1}{4\pi}\frac{\omega_p}{c}E_o \cos k_p l \, dl \, ds = \frac{1}{4\pi}\frac{\omega_p}{c}E_o S_{tr}\int\limits_0^{\lambda_p} \cos k_p l \, dl = 0 , \qquad (A1.9)$$

(here $E_o$ is the amplitude of the photon wave, $S_{tr}$ - the area of torus cross-section, $ds$ is the element of the surface, $dl$ - the element of the circle length, $k_p = \dfrac{\omega_p}{c}$ - the wave-vector absolute value).

It's easy to understand these results: because the ring current changes its direction every half-period, the integral charge is equal to zero.

Therefore, the above model may represent only the non-charged particle. It's clear that the spinning photon in order to become a charge particle must contain only one half-period of wave.

For the solution of this problem we suggest the following **division hypothesis:** *at the moment when the photon begins to roll up, the spontaneous photon division in two half-periods is possible* (see fig.2).

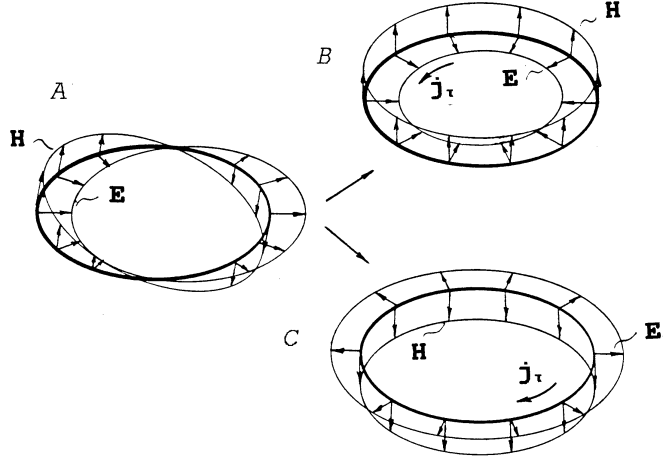

Fig.2. Pair production scheme

It's clear that *the division process corresponds to the process of the particle-antiparticle pair production.* It's proved out by the fact, that the daughter's photons B and C (fig.2) are completely anti-symmetric and can't be transformed one to each other by any transformation of co-ordinates.

It is not difficult to see that *the parts B and C contain the currents of opposite directions.*

We can suppose that *the cause of a photon's division is the mutual repulsion of oppositely directed currents.*

It is interesting that both the circular semi-photon and the circular photon radii must be the same. This fact follows from the momentum conservation low. Really, the spinning photon momentum is equal to:

$$\sigma_p = p_p \cdot r_p = 2m_e c \cdot \frac{\hbar}{2m_e c} = 1\hbar , \qquad (A1.10)$$

In accordance with the momentum conservation low we have (the index "s" refers to the semi-photon):

$$\sigma_s^+ + \sigma_s^- = \sigma_p \tag{A1.11}$$

where $\sigma_s^+, \sigma_s^-$ are the spins of the plus and minus semi-photons. By condition that $\sigma_s^+ = \sigma_s^-$ we obtain:

$$\sigma_s = \frac{1}{2}\sigma_p = \frac{1}{2}\hbar, \tag{A1.12}$$

Since

$$\sigma_s = p_s \cdot r_s, \tag{A1.13}$$

(where $r_s$ is the semi-photon radius, and $p_s = m_e c$ is the inner semi-photon impulse), we have:

$$r_s = \frac{\sigma_s}{p_s} = \frac{1}{2}\frac{\hbar}{m_e c} = \frac{\hbar}{2m_e c} = r_p, \tag{A1.14}$$

So, *the torus size doesn't change after division* and therefore the volumes and areas of the spinning photon and the semi-photon models will be the same: $\Delta\tau_s = \Delta\tau_p$, $S_s = S_{tr}$. The angular rotation velocity (angular frequency) doesn't change also, and $\omega_s = \frac{c}{r_s} = \frac{2m_e c^2}{\hbar} = \omega_p$.

The division hypothesis makes it possible to outline the solution of the some fundamental problems:

1) *the origin of the charge conservation low*: since in nature there are the same numbers of the photon semi-periods of positive and negative directions the sum of the particles charge is equal zero.

2) *the difference between positive and negative charges*: this difference follows from the field and tangent current direction difference of the semi-photons after pair production (by condition that the Pauli-principle is true).

3) *Zitterbewegung*. The results obtained by E.Schroedinger in his well-known articles about the relativistic electron [10] are the most important confirmation for the electron structure model. He showed, that electron has a special inner motion "Zitterbewegung", which has frequency $\omega_z = \frac{2m_e c^2}{\hbar}$, amplitude $r_z = \frac{\hbar}{2m_e c}$, and velocity of light $v_z = c$.

The attempts to explain this motion had not given results (see e.g. [11,12]). But if we suppose, that the electron is the spinning semi-photon, then we receive the simple explanation of Schroedinger's results.

## A1.4. Charge and field mass of spinning semi-photon

We can now calculate the semi-photon charge:

$$q \approx \frac{1}{\pi}\frac{\omega_s}{c} E_o S_s 2\int_0^{\frac{\lambda_s}{4}} \cos k_s l \, dl = \frac{1}{\pi} E_o S_s, \tag{A1.15}$$

Since $S_s = \pi r_s^2$, we obtain:

$$q = E_o r_s^2, \tag{A1.16}$$

To calculate the mass we must calculate first the energy density of the electromagnetic field:

$$\rho_\varepsilon = \frac{1}{8\pi}\left(\vec{E}^2 + \vec{H}^2\right), \quad (A1.17)$$

where $|\vec{E}| = |\vec{H}|$ in Gauss's system. Therefore, equation (3.17) can be written so:

$$\rho_\varepsilon = \frac{1}{4\pi} E^2 , \quad (A1.18)$$

Using the well-known relativistic relationship between a mass and energy densities:

$$\rho_m = \frac{1}{c^2}\rho_\varepsilon , \quad (A1.19)$$

and taking in account (A1.18) we obtain:

$$\rho_m = \frac{1}{4\pi\, c^2} E^2 = \frac{1}{4\pi\, c^2} E_o \cos^2 k_s l , \quad (A1.20)$$

Using (A1.20), we can write for the semi-photon mass (fig.1):

$$m_s = \iint_{S_t, l} \rho_m ds\, dl = \frac{S_s E_o^2}{\pi\, c^2} \int_0^{\frac{\lambda_s}{4}} \cos^2 k_s l\, kl , \quad (A1.21)$$

Calculating the integral, we obtain:

$$m_s = \frac{E_o S_s}{4\omega_s c} = \frac{\pi\, E_o^2 r_s^2}{4\omega_s c} , \quad (A1.22)$$

## A1.5. Relationship between charge and mass of spinning semi-photon

Using equations (A1.16) and (A1.22) we can write:

$$m_s = \frac{\pi\, q^2}{4\omega_s c r_s^2} , \quad (A1.23)$$

or, taking in account that $\omega_s \cdot r_s = c$ we obtain:

$$r_s = \frac{\pi}{2}\frac{q^2}{2m_s c^2} , \quad (A1.24)$$

Put here the values $\omega_s$ and $r_s$, we have:

$$\frac{q^2}{\hbar c} = \frac{2}{\pi} = \alpha_q \approx 0{,}637 , \quad (A1.25)$$

The formula (A1.25) shows, that in presence theory the electric charge is defined only by the constants: $q = \sqrt{\frac{2}{\pi}\hbar c}$; it means that in our theory *free charges don't exist less than this one*. At the same time, this theory doesn't limit the mass value; this fact is according to the experimental data also.

As it is known, the value

$$\frac{e^2}{\hbar c} = \alpha \cong \frac{1}{137} , \quad (A1.26)$$

is the electromagnetic coupling constant. We can suppose that the $\alpha_q$-value distinction from the experimental $\alpha$-value can have as the origin the physical vacuum polarisation [2,13]. As it is known, the virtual particle charges screen the part of the electron charge so that the "bare" electron charge is bigger, than the experimental one. But the calculation problem of the vacuum polarisation is out of the frame of our theory.

Note also that our theory doesn't have the charge infinite problem.

## A1.6. Stability of spinning semi-photon

The solution of this problem is very important for the existence of the above model. In the classical electron theory the forces, which can keep the parts of charge together, had not been found. In our theory such forces exist. The magnetic component of Lorentz's force, which acts in opposite direction to the Coulomb's electric force, appears in the spinning semi-photon.

In fact, as result of the current $\vec{j}_\tau$ and magnetic field $\vec{H}$ interaction, the magnetic force density appears:

$$\vec{f}_M = \frac{1}{c}\left[\vec{j}_\tau \times \vec{H}_s\right], \quad (A1.27)$$

which is directed oppositely to the electric force density. But we cannot calculate their values because we don't know the real field distribution in the circular semi-photon.

## A1.7. Spin of model

Using the data of the torus model we can calculate the spin of the spinning semi-photon:

$$\sigma_s = p_s \cdot r_s = \frac{1}{2}\hbar, \quad (A1.28)$$

Here the difference between bosons and fermions can be well explained if we suppose that *bosons comprise the even number of semi-photons and the fermions comprise the odd number semi-photons.*

## A1.8. Magnetic moment of model

Magnetic moment accordingly with definition is:

$$\mu_s = I \cdot S_I, \quad (A1.29)$$

where $I$ is the electron ring current and $S_I$ is the current ring square.
In our case we have:

$$I = q\frac{\omega_s}{2\pi} = q\frac{1}{2\pi}\frac{2m_e c^2}{\hbar}, \quad (A1.30)$$

$$S_I = \pi\, r_s^2 = \pi\left(\frac{\hbar}{2m_e c}\right)^2, \quad (A1.31)$$

Using these formulae, we find:

$$\mu_s = \frac{1}{2}\frac{q\hbar}{2m_e}, \qquad (A1.32)$$

If we put $q = e$, the value (A1.32) is equal to half of the experimental value of the magnetic momentum of the electron. Taking into account Thomas's precision we obtain the experimental value of the electron magnetic momentum.

The above model, of course, cannot give the answers on all of the questions but completes the exact theory.

## Conclusion

The present interpretation doesn't contradict to the quantum mechanics, but explains well its results (for instance, the electrical charge generation, the existing of elementary charge, the charge universality, the appearance and difference between the negative and positive charges, the electric charge conservation law, particle spin, difference between bosons and fermions, the sense of bispinor forms of Dirac's equation, and many other).

Note that the above theory can be developed in the other areas of the quantum theory.